\documentclass[sigconf]{acmart}
\AtBeginDocument{%
  }

\setcopyright{acmlicensed}
\copyrightyear{2026}
\acmYear{2026}
\setcopyright{cc}
\setcctype{by}
\acmConference[ICPP '26]{Proceedings of the 55th International Conference on Parallel Processing}{September 28-October 01, 2026}{Singapore, Singapore}
\acmBooktitle{Proceedings of the 55th International Conference on Parallel Processing (ICPP '26), September 28-October 01, 2026, Singapore, Singapore}
\acmDOI{10.1145/3832810.3832862}
\acmISBN{979-8-4007-2657-6/2026/09}

\begin{document}

\title{NeuroPrefetcher: Storage-Aware Sparse LLM Inference via Delta Prefetching}

\author{Nobel Dhar}
\email{ndhar@students.kennesaw.edu}
\orcid{0009-0000-5467-7082}
\affiliation{%
  \institution{Kennesaw State University}
  \city{Marietta}
  \country{USA}
}

\author{Md Romyull Islam}
\email{mislam22@students.kennesaw.edu}
\orcid{0000-0003-0550-2248}
\affiliation{%
  \institution{Kennesaw State University}
  \city{Marietta}
  \country{USA}
  \email{}
  \orcid{}
}

\author{Xuechen Zhang}
\email{xzhang54@kennesaw.edu}
\orcid{0000-0002-3730-8901}
\affiliation{%
  \institution{Kennesaw State University}
  \city{Marietta}
  \country{USA}
}

\author{Gongjin Sun}
\email{gongjin.s@samsung.com}
\orcid{0009-0002-5420-1361}
\affiliation{%
  \institution{Memory Solutions Lab, Samsung Semiconductor, Inc.}
  \city{San Jose, California}
  \country{USA}
}

\author{Sahidul Islam}
\email{sislam19@kennesaw.edu}
\orcid{0000-0002-4488-8182}
\affiliation{%
  \institution{Kennesaw State University}
  \city{Marietta}
  \country{USA}
}

\author{Bobin Deng}
\email{bdeng2@kennesaw.edu}
\orcid{0000-0001-8361-9025}
\affiliation{%
  \institution{Kennesaw State University}
  \city{Marietta}
  \country{USA}
}

\author{Kun Suo}
\authornote{Corresponding author.}
\email{ksuo@kennesaw.edu}
\orcid{0000-0001-8562-0492}
\affiliation{%
  \institution{Kennesaw State University}
  \city{Marietta}
  \country{USA}
}


\begin{abstract}
Deploying large language models on edge devices is increasingly limited by a
widening gap between model size and available memory. Existing approaches such
as quantization, smaller models, and offloading can raise the effective memory
limit, but they still assume that the model can be compressed or partitioned to
fit within some budget. We target the harder model-exceeds-memory
setting, in which the model remains larger than resident memory throughout
execution and storage becomes an active source of weights on the critical path.
We observe that MLP activity during autoregressive decoding has strong
temporal locality: approximately 82--85\% of active neurons persist
from one token to the next. This means that most sparse weights needed for the
current token are already resident, and only the newly needed rows must be fetched from storage. We
present NeuroPrefetcher, a storage-backed LLM inference system that
exploits this property through predictive delta prefetching. After
layer~0, a single GPU-resident predictor, occupying 2.86\% of base model
parameters, predicts sparse activity for all downstream MLP layers in one
forward pass. The runtime compares these predictions against resident GPU
buffers and issues application-scheduled NVMe reads only for incoming delta
rows, replacing reactive operating-system demand paging with explicit,
model-aware weight movement.
On real unified-memory edge hardware, NeuroPrefetcher achieves
7.9--12.0$\times$ speedup over llama.cpp across constrained memory
budgets.
\end{abstract}




\maketitle

\section{Introduction}

Large language models (LLMs) now underpin many language applications, but their
growing capability comes with a growing systems cost. As models scale,
inference demands more memory capacity, more memory bandwidth, and stronger
accelerator support~\cite{openai2024gpt4technicalreport,mistral}.
This tension is especially severe on edge platforms, where memory growth lags
far behind model growth. Over the past five years, model size has increased by
roughly $10\times$ per generation, while edge DRAM capacity has increased by
only about $2\times$~\cite{scalinglaws,dhar2024acmse}. The result is a widening
\textbf{model--memory gap}: increasingly capable models do not merely run slowly
on edge devices; many no longer fit at all.

\begin{figure*}[t]
    \centering
    \includegraphics[width=0.9\textwidth]{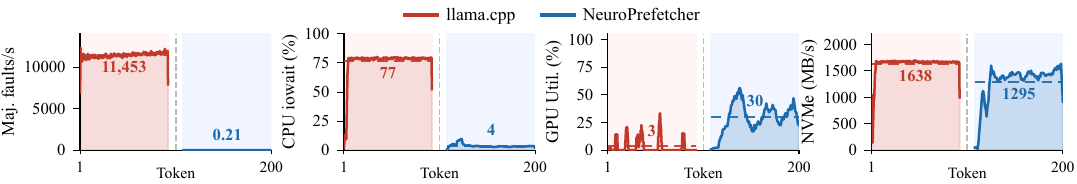}
   \caption{System-level profile during LLM inference on Jetson AGX Orin.}
   \Description{Four time-series panels comparing llama.cpp and NeuroPrefetcher during decoding, showing major page faults per second, CPU iowait percentage, GPU utilization percentage, and NVMe bandwidth.}
  \label{fig:motivation}
\end{figure*}

Most existing techniques respond by moving the model back under a memory
limit. Quantization reduces the number of bytes per weight, and methods
such as GPTQ and AWQ can compress a 7B model from 13.5\,GiB to roughly
4\,GiB at 4-bit precision~\cite{gptq,awq}. Smaller models such as Phi-3 and
MobileLLM are designed to fit edge devices, but they reduce model capacity to
meet a fixed memory budget~\cite{phi3,mobilellm}. Offloading systems such as
llama.cpp and FlexGen keep the original model but place weights across GPU
memory, CPU memory, and storage~\cite{llamacpp,flexgen}. All three approaches
assume the model can be made to fit within a memory budget. They are not
designed for the harder setting in which the model remains larger than
resident memory throughout execution. This setting arises not only when a
model outgrows a device but whenever the memory available to inference falls
below the model footprint, as happens when the LLM shares an edge platform
with other workloads or runs on a smaller device. In this setting, storage
becomes an \textbf{active weight tier}, a part of the memory hierarchy that
must directly supply weights during inference, and performance depends not
only on tensor computation but on how efficiently the system moves weights
between storage and compute.

Existing systems handle that movement mostly reactively. For example,
when the model exceeds memory, llama.cpp~\cite{llamacpp} falls back to
memory-mapped weights and OS demand paging. Under demand paging, missing
weights are loaded only after execution touches them.
Figure~\ref{fig:motivation} profiles this path on a Jetson AGX Orin running
Mistral-7B at \texttt{mem=14G}. During decoding, the system generates 11{,}453
major page faults per second. The CPU spends 77\% of its time blocked on I/O,
while GPU averages only 3\% utilization. Inference in this setting is limited
not by arithmetic throughput but by \textbf{reactive movement of model
weights}.

One way to reduce weight movement is \textbf{activation sparsity}. For a given
token, many MLP neurons contribute little to the output and need not
participate in the MLP computation. Prior work has shown that ReLU-based
models expose more than 90\% natural
sparsity~\cite{dejavu,relustrikesback}. Modern SwiGLU models such as Mistral
and Llama-2/3 do not produce the same exact-zero pattern, but a large fraction
of their MLP activations can still be removed through magnitude thresholding
or learned sparsification with limited quality
loss~\cite{teal,dhar2024ipccc,cats,turbosparse}. In a storage-backed runtime,
sparsity reduces not only computation but also the weight data that must be
read from storage. However, most sparse runtimes reduce the weight movement of
a single decoding step. They do not directly minimize the \emph{new} storage
traffic introduced at each successive token.

The missing opportunity is \textbf{temporal locality} across tokens. In
autoregressive decoding, adjacent tokens often activate similar subsets of MLP
neurons. Our key observation is that active MLP neurons persist strongly from
one token to the next, so most of the current token's sparse weights are
already resident and only the \textbf{delta}, the newly needed rows that were
not already resident, must be fetched. Exploiting this locality requires two
capabilities at once. First, the runtime must track which weight rows are
already resident for each layer. Second, it must know the next token's active
set early enough to fetch the delta before sparse execution reaches that
layer. Existing systems do not provide both. PowerInfer places hot neurons
near compute using offline profiling, but targets the GPU-to-CPU-memory tier
rather than NVMe-backed cold weights~\cite{powerinfer}. LLM in a Flash retains
recently activated neurons with a backward-looking window, while Neuralink
co-locates co-activated neurons in flash to make sparse reads more
sequential~\cite{llminaflash,neuralink}. These techniques reduce computation,
placement cost, reuse-miss cost, or sparse-read cost, but none provides a
lookahead mechanism that predicts the next sparse working set and transfers
only the incremental rows needed for execution.

\begin{figure*}[t]
    \centering
    \includegraphics[width=0.88\textwidth]{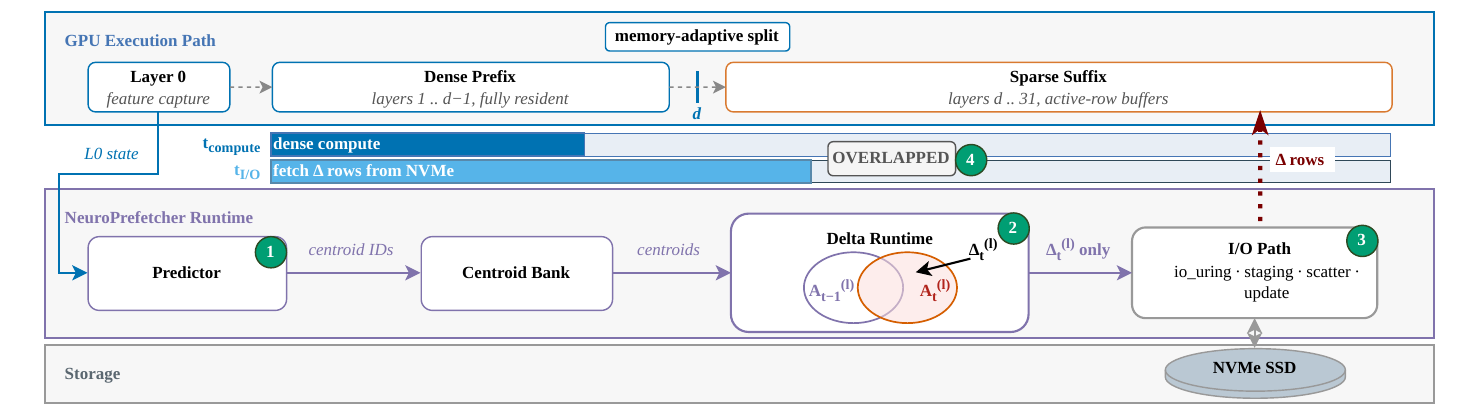}
    \caption{High-level NeuroPrefetcher architecture with predictor-guided weight prefetching and sparse MLP execution.}
    \Description{Block diagram showing the GPU execution path with a dense prefix and sparse suffix, the NeuroPrefetcher runtime with predictor, centroid bank, delta runtime, and I/O path, and the NVMe storage tier supplying delta rows.}
    \label{fig:system-overview}
\end{figure*}

This paper addresses these gaps with \textbf{NeuroPrefetcher}, a
storage-backed LLM inference system built around predictive delta
prefetching. Predictive delta prefetching means that the system predicts
which sparse weights the next token will need, compares that predicted set
against the weights already resident, and fetches only the difference in
advance. First, NeuroPrefetcher predicts sparsity early enough to drive
lookahead I/O. After layer~0, a single GPU-resident predictor uses early-token
features to predict active-neuron masks for all downstream MLP layers in one
forward pass, making the sparse-layer I/O schedule available before sparse
execution begins, unlike per-layer predictors such as DejaVu that reveal a
layer's active set only when execution reaches it~\cite{dejavu}. Second, the
runtime turns predictions into movement decisions, comparing each layer's
predicted active set with the rows already resident and fetching only the
incoming delta from NVMe while the GPU computes on resident layers. Because
direct I/O on Jetson's unified-memory platform can bypass CPU cache coherence
and expose stale data to the GPU, the runtime routes sparse rows through a
cache-coherent staging path before GPU execution. Third, a memory-adaptive
\textbf{dense/sparse split} divides GPU memory between a dense prefix with
fully resident early layers and a sparse suffix holding active-row buffers,
trading dense layers for sparse ones as memory tightens. The right panels of
Figure~\ref{fig:motivation} show the consequence on the same hardware: major
page faults fall to 0.21 per second, CPU iowait drops to 4\%, and GPU
utilization rises to 30\%.

We evaluate NeuroPrefetcher on a Jetson AGX Orin with 32\,GiB unified memory
and a Samsung 990 PRO NVMe SSD, running Mistral-7B-v0.1 and Llama-3-8B under
constrained memory budgets. In the model-exceeds-memory regime,
NeuroPrefetcher achieves 7.9--12.0$\times$ speedup over llama.cpp, reaching
3.22\,tok/s against 0.34\,tok/s at \texttt{mem=14G}. In a capability study
across twenty production inference systems, twelve fail before generation and
four lack a deployable release for this platform, exposing weight-transfer
allocation paths that assume resident weights and discrete-GPU execution
assumptions that do not hold on unified-memory edge hardware. NeuroPrefetcher
is the fastest of the four systems that produce output and the only distinct
engine designed for sparse storage-backed execution in this regime. In summary, this paper makes the following contributions:

\begin{itemize}
\item We introduce a prefetching predictor that exposes the sparse-layer
I/O plan after layer~0. The runtime fetches only the rows the current
token newly requires, reusing 82--85\% of the previous token's
active set. This predictor preserves 92--96\% of dense accuracy
without per-task retraining.

\item We build an explicit NVMe-to-GPU sparse-row path that replaces reactive
page faults with application-scheduled weight movement. The runtime pairs
this path with a memory-adaptive dense/sparse split. As memory tightens,
NeuroPrefetcher trades dense layers for sparse ones rather than falling
back to much larger whole-layer NVMe reads.

\item We evaluate twenty production LLM inference systems on unified-memory
edge hardware. NeuroPrefetcher is the fastest among the operational systems
and the only distinct engine designed for sparse storage-backed execution
in this regime.
\end{itemize}

\section{Related Work}
\label{sec:related}

\noindent\textbf{Reducing Model Footprint.}
Quantization, pruning, distillation, and compact model design all reduce the
memory footprint before deployment. Quantization lowers the bytes per weight;
methods such as GPTQ~\cite{gptq} and AWQ~\cite{awq} compress 7B models to
roughly 4\,GiB at 4-bit precision, and llama.cpp~\cite{llamacpp} deploys such
formats on commodity hardware. Pruning removes weights judged
unimportant~\cite{pruning}. Distillation and compact model design train
smaller models to fit a fixed budget~\cite{hinton2015distilling}. These
techniques move the fit point, but they still assume the deployed model can be
made resident, and model size continues to grow faster than edge memory
capacity. NeuroPrefetcher addresses the regime after that assumption breaks,
where the full model remains on storage and the runtime reduces how much
weight data moves per token. Quantization is orthogonal to NeuroPrefetcher.
Lowering weight precision leaves the neuron count and activation structure
unchanged, so NeuroPrefetcher remains applicable to quantized models.
Quantization is the natural first response to a memory deficit, and
NeuroPrefetcher targets the deployments where even the quantized model
exceeds memory.

\begin{figure*}[t]
    \centering
    \includegraphics[width=0.88\textwidth]{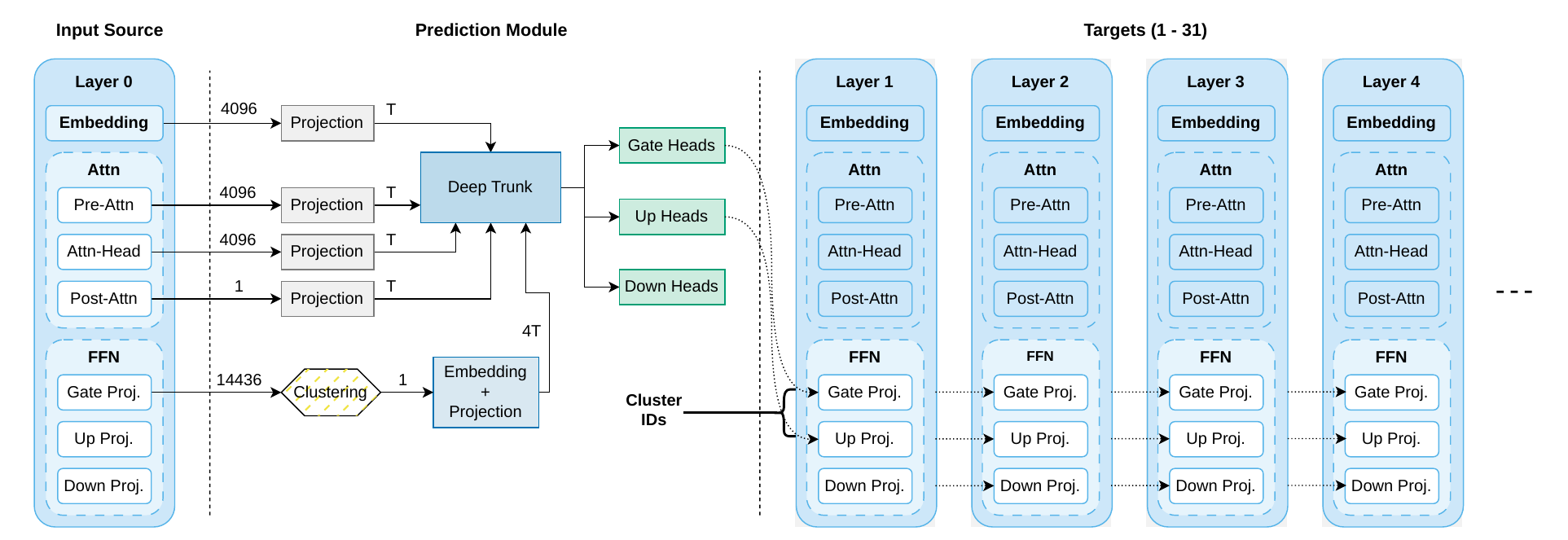}
    \caption{One-shot shared predictor for gate and up centroid IDs.}
    \Description{Diagram of the shared predictor. Layer 0 features feed four projection blocks and a centroid embedding, a deep residual trunk combines them, and separate gate and up heads emit one centroid ID per downstream layer.}
    \label{fig:predictor-architecture}
\end{figure*}

\noindent\textbf{Activation Sparsity.}
DejaVu~\cite{dejavu} predicts contextual sparsity in attention heads and MLP
neurons using per-layer predictors and shows that sparse execution can improve
wall-clock performance. Its predictor reveals a layer's active set only when
execution reaches that layer, which is too late for storage-backed lookahead
prefetching; NeuroPrefetcher instead uses one predictor after layer~0 to
expose the sparse-layer I/O plan before sparse execution begins. Other work
exploits natural or enforced sparsity. ReLU-based models expose large
exact-zero activation patterns~\cite{relustrikesback}, and
SparseInfer~\cite{SparseInfer} uses this structure for sparse execution, while
thresholding and learned sparsification remove many low-magnitude activations
from SwiGLU models with limited quality
loss~\cite{dhar2024ipccc,cats,turbosparse}. These methods mainly reduce
computation. NeuroPrefetcher uses sparsity as an I/O signal, predicting which
MLP rows should cross the NVMe boundary and avoiding reloads of rows that
remain resident across tokens.

\noindent\textbf{Memory and Storage Systems.}
Offloading systems such as FlexGen~\cite{flexgen} and llama.cpp~\cite{llamacpp}
place weights across accelerator memory, host memory, and storage. They raise
the effective memory limit, but storage remains reactive: weights move when
execution reaches them, through streaming or memory-mapped paging, whereas
NeuroPrefetcher schedules storage movement before sparse execution begins.
PowerInfer~\cite{powerinfer} profiles neurons offline, keeps hot neurons on
the GPU, and executes cold neurons on the CPU. This assumes CPU DRAM is a
separate capacity tier; on unified-memory edge devices, host offload creates
no new capacity, and cold weights must be served from NVMe. PowerInfer also
uses per-layer prediction, while NeuroPrefetcher uses one-shot prediction to
drive lookahead storage prefetching. LLM in a Flash~\cite{llminaflash} retains
recently activated neurons with a backward-looking window, and
Neuralink~\cite{neuralink} co-locates co-activated neurons in flash to make
sparse reads more sequential. These systems improve storage-backed sparse
access, but none maintains a per-layer resident active set updated by
token-to-token set difference, which is the mechanism that lets
NeuroPrefetcher fetch only incoming active rows.
PRESERVE~\cite{PRESERVE} prefetches whole-layer weights and
KV-cache blocks in distributed serving, and ShadowKV~\cite{sun2025shadowkv}
compresses KV-cache state for long-context workloads. These systems target
distributed transfer or KV-cache pressure, not sparse MLP weight movement
from NVMe on memory-constrained edge devices.

\section{Design of NeuroPrefetcher}

NeuroPrefetcher combines three pieces: an \emph{offline preparation}
stage that converts dense MLP activations into sparse centroids, a
\emph{shared predictor} that selects those centroids at runtime, and a
\emph{storage-backed execution engine} that fetches only the newly
needed weights. This section describes those pieces in the same order
in which they are used.

\subsection{System Overview}
\label{subsec:overview}
 \looseness=-1 
NeuroPrefetcher splits the model's $N$ transformer layers into two
regions. The first $d$ layers are \emph{dense}: their MLP weights stay
fully resident in GPU memory. The remaining $N{-}d$ layers are
\emph{sparse}: each layer keeps only a fixed-capacity GPU buffer of
$K_\mathrm{max}$ neuron rows, while the full sparse weight set is stored on
NVMe.
\looseness=-1 
Some weights remain resident regardless of this split: the token
embedding table, language modeling head, final normalization,
attention projections for all layers, per-layer normalization
parameters, the shared predictor, and the centroid table. We choose $d$
with an automated sweep that finds the largest dense prefix that fits
within the memory budget. Section~\ref{sec:ratio} shows how this
choice affects throughput and when the engine transitions to a
secondary operating mode.

Figure~\ref{fig:system-overview} shows the per-token execution flow.
Step \ding{172}: the engine runs layer~0 densely and extracts the features needed
by the shared predictor. Step \ding{173}: the predictor emits one centroid ID
for the gate projection and one centroid ID for the up projection of
each later layer. These IDs define the predicted active neuron set for
each sparse layer; comparing that set with the rows already resident in
the layer's GPU buffer yields the \emph{delta}, i.e., the newly needed
rows that must be fetched for the current token. Step \ding{174}: a background thread issues batched NVMe reads for the delta
rows. Because each neuron's gate, up, and down rows are stored
contiguously, one storage read retrieves all three projections, and
consecutive neuron IDs can be coalesced into larger ranges. Step \ding{175}:
the completed reads land in a pinned staging buffer, and a CUDA scatter
kernel places each row into its assigned position in the per-layer GPU
buffer. While this I/O path runs, the main thread continues executing
the dense prefix on the GPU. Finally, once the delta rows are ready,
the engine executes the sparse suffix using permanently resident
attention weights and the refreshed MLP row buffers. At the first token
of a sequence, the engine loads the full predicted active set for each
sparse layer; after that, it updates only the delta.

\subsection{Offline Sparse Centroid Preparation}
\label{subsec:centroid_preparation}

NeuroPrefetcher needs a compact runtime representation of sparse MLP activation
patterns. Directly predicting one binary state for every MLP neuron in every
downstream layer would make the predictor large and difficult to train. 
Instead, we first build a compact vocabulary of sparse activation
patterns offline, then train the predictor to choose from that
vocabulary at runtime. This preparation stage has two steps:
thresholding and clustering.

\subsubsection{Activation Thresholding for Sparsity Enforcement}
\label{subsec:activation_thresholding}

We begin by inducing sparsity in the MLP activations.
Modern LLMs often use smooth activations such as GeLU or SiLU, which do not
naturally produce exact zeros. In our measurements, many MLP neurons still
produce very small values across tokens. These low-magnitude neurons contribute
little to the output but still require weight movement and matrix-vector work.
MLP blocks contain most model parameters, so they are the main target for
storage-backed sparsity. We leave attention blocks unchanged because changing
attention tends to have a larger effect on model quality.

Following prior work~\cite{dhar2024ipccc}, we enforce sparsity with a
per-layer percentile threshold. For layer $l$, we collect activations
$\mathbf{A}_{i,l}=f_l(\mathbf{x}_i)$ over tokens $\mathbf{x}_i$ and compute
$T_{l,\alpha}$ as the $\alpha$-th percentile of the magnitude distribution
$|\mathbf{A}_{i,l}|$. We apply
\begin{equation}
\mathbf{A}_{i,l}^{\mathrm{sparse}}[j] =
\begin{cases}
0 & \text{if } \left| \mathbf{A}_{i,l}[j] \right| < T_{l,\alpha}, \\
\mathbf{A}_{i,l}[j] & \text{otherwise}.
\end{cases}
\end{equation}
The sparsity level $\alpha$ controls the active-neuron budget. Higher
$\alpha$ removes more low-magnitude neurons and reduces storage traffic, while
lower $\alpha$ keeps more neurons and preserves quality. This step turns each
dense MLP activation vector into a sparse activation pattern without retraining.

\subsubsection{Clustering of Sparse Activation Patterns}
\label{subsec:clustering}

\begin{figure}[h]
  \centering
  \includegraphics[width=\columnwidth]{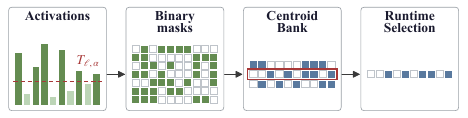}
  \caption{Offline sparse centroid preparation.}
  \label{fig:clustering}
\end{figure}

Thresholding alone is not enough, because the prediction problem remains too large, as established above. We therefore cluster
thresholded activation masks and predict only \emph{centroid IDs}.

After thresholding, each MLP activation vector is converted into a
binary mask whose ones mark active neurons. Across tokens and layers,
these masks are numerous and highly variable. We reduce that
variability by grouping masks that activate similar neuron sets and by
representing each group with a sparse binary centroid.
Figure~\ref{fig:clustering} illustrates this process. The number of
clusters is a tunable parameter: more clusters preserve finer-grained
patterns, but they also make the predictor's classification task
harder.

We build on earlier work on clustering high-dimensional binary
activations~\cite{ndhar3}. Each centroid is formed by aggregating the
member masks of a cluster and then pruning the aggregate back to the
target sparsity level, so the resulting centroid remains sparse by
construction. \looseness=-1 This clustering step reduces the runtime prediction space
from thousands of neuron decisions to one centroid choice per layer,
while still preserving the sparse structure needed for execution.

\subsection{Shared Activation Predictor}
\label{subsec:predictor}

Once the centroid vocabulary is generated, we train a lightweight predictor
that maps early-token features to centroid IDs for later MLP layers.
The predictor runs once per token and outputs $2 \times 31 = 62$
centroid identifiers: one for the gate projection and one for the up
projection of each of layers 1 through 31. We leave down projections
dense in all layers, because their 4{,}096-neuron dimension is small
enough that sparsifying them noticeably hurts output quality.
Figure~\ref{fig:predictor-architecture} shows the architecture.

\begin{table}[h]
\centering
\small
\caption{Layer~0 features consumed by the shared predictor.}
\label{tab:predictor_inputs}
\begin{tabular}{lc}
\toprule
\textbf{Feature} & \textbf{Input form} \\
\midrule
Token embedding & 4096-d vector \\
Pre-attention LayerNorm & 4096-d vector \\
Attention-head sketch & 128-d vector \\
Post-attention RMS norm & 1 scalar \\
Gate projection structure & centroid ID \\
\bottomrule
\end{tabular}
\end{table}

\subsubsection{Input Features}

The predictor uses five features extracted from layer~0, summarized in
Table~\ref{tab:predictor_inputs}. Four of them, the token embedding,
pre-attention layer normalization output, attention sketch, and
post-attention residual norm, describe the token's early semantic and
contextual state. The fifth feature captures \emph{activation
structure}: we take the layer~0 gate activations, map them to their
nearest centroid, and pass only that centroid ID. This compact code is
much cheaper than feeding the full 14{,}336-dimensional vector, while
still preserving information about which neurons are likely to activate
in deeper layers.

\subsubsection{Architecture}

The predictor has three stages as shown in Figure~\ref{fig:predictor-architecture}. 
First, each of the four continuous
semantic features pass through its own projection block consisting of a linear layer, LayerNorm, SiLU, and dropout, producing a $T{=}512$-dimensional embedding. Their concatenation forms a $4T$-dimensional semantic
representation. In parallel, the layer~0 gate centroid ID is processed
by a learned embedding and a small projection network to produce a
second $4T$-dimensional vector that captures activation structure. The
two are concatenated into an $8T$-dimensional input.

Second, this input is processed by a residual MLP trunk. Each block
compresses the representation to $4T$, applies LayerNorm, SiLU, and
dropout, then expands back to $8T$ with a residual connection. This
trunk combines semantic and structural signals into a shared feature
space.

Third, two groups of 31 independent linear heads predict centroid
logits: one group for gate projections and one for up projections
across layers 1 through 31. Formally,
$\mathbf{G} \in \mathbb{R}^{31 \times K}$ and
$\mathbf{U} \in \mathbb{R}^{31 \times K}$ are the gate and up logits
over the centroid vocabulary of size $K$. At inference time, each head
selects its highest-probability centroid.

\subsubsection{Training}

The predictor is trained with cross-entropy over gate and up centroid IDs:
\[
\mathcal{L}
=
\mathcal{L}_{\mathrm{CE}}^{\mathrm{gate}}
+
\mathcal{L}_{\mathrm{CE}}^{\mathrm{up}} .
\]
Training uses AdamW, distributed data parallelism, a batch size of 2{,}048
tokens per GPU, and gradient clipping at 1.0. We use $T=512$ and a four-block
residual trunk, and we evaluate centroid counts
$K=512$. The trained predictor is fixed at runtime and runs once
per generated token.

\begin{table}[t]
\centering
\caption{Predictor parameter breakdown.
Sizes are FP16.}
\label{tab:predictor_cost}
\begin{tabular}{lcc}
\toprule
\textbf{Component} & \textbf{Params} & \textbf{Size (MB)} \\
\midrule
Input projections   & 4.3M   & 8.6 \\
Centroid embedding  & 5.2M   & 10.4 \\
Residual trunk      & 67.1M  & 134.2 \\
Output heads        & 130.2M & 260.4 \\
\midrule
Total               & 206.8M & 413.6 \\
\bottomrule
\end{tabular}
\end{table}

\subsubsection{Predictor Cost}
The predictor contains 206.8M parameters, or 2.86\% of the base model, and
occupies 414\,MB in FP16. It is among the always resident weights and all reported speedups are net of this
cost. Table~\ref{tab:predictor_cost} breaks down the parameter count and the
scaling behavior of each component. The output heads dominate and grow with
the number of layers, the residual trunk is fixed, and no component grows
with the MLP intermediate dimension, which is the dimension that grows
fastest across model families.

\subsection{Predictive Delta Prefetching}
\label{subsec:delta}

The shared predictor tells the runtime which neurons are expected to be
active in each sparse layer. The runtime then updates each layer's GPU
buffer incrementally instead of reloading the full active set every
token. Each sparse layer maintains a fixed-capacity GPU buffer of
$K_\mathrm{max}$ rows, and the runtime transfers only the \emph{delta}:
the rows that are newly active for the current token.

\begin{figure}[t]
  \centering
  \includegraphics[width=0.88\columnwidth]{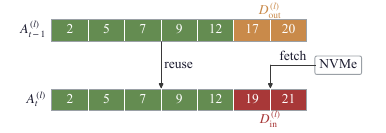}
  \caption{Token-to-token delta update for one sparse MLP layer.}
  \label{fig:delta}
\end{figure}

\looseness=-1 
The predictor produces the active set
$A_t^{(l)}$ for layer $l$ at token $t$ by intersecting the gate and up centroid
masks. This is sufficient for a SwiGLU MLP because an intermediate neuron
contributes only when both its gate and up terms are active. The same active set
therefore indexes all three projection buffers. For each active neuron $i$, the
runtime uses row $i$ of the gate, up, and $\mathrm{down}^{\mathsf T}$
projections.
\looseness=-1 
Reloading the full active set every token would repeat most of the same NVMe
traffic. NeuroPrefetcher instead updates each resident buffer by set difference:
\begin{align}
    D_\mathrm{in}^{(l)}  &= A_t^{(l)} \setminus A_{t-1}^{(l)}, \\
    D_\mathrm{out}^{(l)} &= A_{t-1}^{(l)} \setminus A_t^{(l)} .
\end{align}
Rows in $D_\mathrm{in}^{(l)}$ are incoming active rows and are fetched from
NVMe. Rows in $D_\mathrm{out}^{(l)}$ leave the active set and release their
buffer slots. Each active set has fixed size $K_\mathrm{max}$, so incoming rows
closely match outgoing rows. The runtime reuses freed slots and avoids buffer
compaction. Figure~\ref{fig:delta} illustrates the buffer transition between two consecutive tokens.

A per-layer slot map $S^{(l)} \in \mathbb{Z}^{M}$ tracks which neurons are
resident, where $M$ is the intermediate dimension. Entry $S^{(l)}[n]$ stores the
buffer slot for neuron $n$, or $-1$ when the neuron is not resident. At each
token, runtime clears slot-map entries for outgoing rows, assigns freed
slots to incoming rows, and updates the incoming entries. The sparse MLP kernel
uses $S^{(l)}$ to locate each active row. Row order inside the buffer does not
matter. Correctness depends on row presence and slot address, not sorted neuron
order.

The runtime schedules NVMe reads only for incoming active rows. Read ranges
across sparse layers are batched into one token-level submission. The first
token has no previous active set, so the runtime initializes each sparse layer
by fetching its full predicted active set and populating the slot map. All later
tokens use the delta path. After the cold-start token, the ratio
$|D_\mathrm{in}^{(l)}|/K_\mathrm{max}$ drops quickly and stabilizes in steady state.

\subsection{I/O Path}
\label{subsec:io}

The delta runtime produces a token-level list of incoming active rows. This
section describes how the I/O path moves those rows from NVMe into resident
active-row buffers. The path combines one-time sparse weight layout preparation
with runtime coalescing, asynchronous reads, cache-coherent staging, and fused
GPU scatter. The core mechanism is a model-aware sparse-row transfer path.
Jetson adds a unified-memory coherence constraint, which the runtime handles
through mapped pinned staging.

We prepare a \emph{neuron-centric weight layout} for sparse MLP weights.
Standard checkpoint formats store the gate, up, and $\mathrm{down}^{\mathsf T}$ projection matrices as
separate tensors, so fetching one active intermediate neuron requires separate
ranges from each projection matrix. We instead repack the sparse MLP weights
once, offline, into a contiguous layer-major file. Each 24.5\,KiB record
co-locates the gate, up, and $\mathrm{down}^{\mathsf T}$ projection rows for one
intermediate neuron. The down projection is transposed during offline
preprocessing, so all three rows share the same row-major shape and stride. This
layout aligns storage order with the sparse execution unit. A single record read
returns all projection rows needed for one active neuron, and consecutive
active-neuron IDs map to consecutive byte ranges within a layer.

\begin{table}[t]
\centering
\small
\caption{Memory levels and measured CUDA availability.}
\label{tab:memory_levels}
\begin{tabular}{ccc}
\toprule
Kernel \texttt{mem=} & CUDA total (GiB) & CUDA free (GiB) \\
\midrule
11G & 10.69 & 9.02 \\
12G & 11.67 & 9.99 \\
13G & 12.65 & 11.02 \\
14G & 13.64 & 11.94 \\
15G & 14.62 & 13.19 \\
16G & 15.60 & 13.88 \\
17G & 16.56 & 14.81 \\
\bottomrule
\end{tabular}
\end{table}

At runtime, NeuroPrefetcher sorts incoming active-neuron IDs and coalesces
adjacent IDs into contiguous read ranges. The I/O thread submits these ranges
through Linux \texttt{io\_uring} with the file opened using \texttt{O\_DIRECT}.
The runtime maintains an \texttt{io\_uring} queue depth of 256, which holds
enough reads in flight to saturate the drive for this access pattern while the
main thread executes the dense prefix. Deeper queues yield no additional
bandwidth for coalesced record reads but consume proportionally more staging
memory, since each in flight request holds its own read buffer, and increase
tail latency. Completed NVMe reads
first land in a host-side staging buffer before CUDA scatter places them into
resident active-row buffers. This staging step is required for correctness on
Jetson AGX Orin. The CPU and GPU share physical memory, but direct I/O can write
through a path that is not immediately visible to CUDA kernels. Directly
scattering from an ordinary host buffer can expose stale data. The runtime
therefore uses page-aligned host staging memory registered with
\texttt{cudaHostRegisterMapped}. CUDA kernels read the staged records through a
coherent mapped path without an explicit \texttt{cudaMemcpy}.

\looseness=-1

Rows arrive in staging memory as interleaved gate/up/down$^{\mathsf T}$
records. GPU execution uses three projection-specific resident buffers:
$\mathrm{sparse\_gate}$, $\mathrm{sparse\_up}$, and
$\mathrm{sparse\_down}^{\mathsf T}$. A fused CUDA kernel,
\texttt{launch\_deinterleave\_scatter}, reads each staged record and writes its
three rows into the slot assigned by the Delta Runtime. Fusing deinterleave and
scatter removes an intermediate device buffer and preserves row-major access for
the next sparse MLP launch. The scatter runs on a dedicated CUDA stream, so its
tail can overlap with dense-prefix compute.
The staging buffer is capacity-bounded and sized for the steady-state token
delta rather than the largest possible cold start. If a token delta exceeds
staging capacity, the runtime partitions the read list into independent
submit-and-scatter chunks. Large deltas reduce overlap, but they do not require
a new allocation and do not terminate execution. Beyond this platform, the coherence staging step is the only Jetson-specific component.
The predictor, delta runtime, neuron-centric layout, and asynchronous read
submission are independent of the memory organization.

\begin{table}[h]
\centering
\small
\caption{Capability results at a 14\,GiB memory cap with 11.9\,GiB CUDA-available memory.}
\label{tab:capability}
\begin{tabular}{@{}p{0.28\columnwidth}p{0.68\columnwidth}@{}}
\toprule
\textbf{Outcome} & \textbf{Systems} \\
\midrule
operational & NeuroPrefetcher (3.22 tok/s), llama.cpp FP16 (0.34), FlexGen all-disk (0.25), Ollama (0.07) \\
\midrule
alloc.\ failure & HF Transformers, HF Transformers (disk), ZeRO-Inference, CATS, TEAL, SGLang \\
\midrule
unified-memory mismatch & vLLM, TensorRT-LLM, MLC-LLM, PowerInfer, Candle, DejaVu \\
\midrule
no ARM/JetPack release & LLM in a Flash, LMDeploy, TGI, HF Optimum + ONNX GPU \\
\bottomrule
\end{tabular}
\end{table}

\section{Evaluation}

\subsection{Experimental Setup}
\label{sec:setup}

\noindent
\textbf{Hardware.}
We run all throughput, latency, and I/O experiments on an NVIDIA Jetson AGX
Orin developer kit. The board has 32\,GiB of unified LPDDR5 memory shared by a
12-core Arm Cortex-A78AE CPU and an Ampere GPU with 2048 CUDA cores
(SM~8.7). Storage is a Samsung 990 PRO NVMe SSD connected through M.2. The SSD
is rated for 7.4\,GB/s sequential reads. Offline centroid preparation and predictor training are performed on a server
with 8 NVIDIA A100 80\,GB GPUs.

On edge platforms, the memory available to inference is a deployment parameter
rather than a hardware constant. Perception pipelines, application logic, and
system services consume a substantial and varying share of device memory.
Smaller devices in the same product family offer even less physical memory to
begin with. To study different memory budgets on the edge device, we therefore
constrain total system memory using the Linux kernel memory limit boot
parameter. We test seven levels, from 11\,GiB to 17\,GiB in 1\,GiB steps.
After the operating system, display server, and CUDA runtime consume their
fixed share, the available CUDA memory for our process ranges from 9.0\,GiB to
14.8\,GiB. Table~\ref{tab:memory_levels} reports the measured memory levels.
On Jetson's unified memory platform, this value reflects the resident memory
available to our process for both CPU and GPU allocations.

\noindent
\textbf{Models and sparsity.}
Throughput, latency, and I/O experiments use Mistral-7B-v0.1~\cite{mistral} in
FP16. The model has 7.24 billion parameters, a 13.49\,GiB weight footprint, and
32 transformer layers with SwiGLU MLP blocks. Its hidden size is 4096 and its
intermediate size is 14{,}336. The quality study also evaluates Llama-3-8B~\cite{llama3} using the same sparse execution methodology.

We enforce activation sparsity by retaining only the predicted active MLP
neurons for each token. The active count is controlled by
$K_{\mathrm{max}}$, the maximum number of active intermediate neurons kept per
sparse layer. We evaluate three budgets:
\begin{itemize}
    \item $K_{\mathrm{max}} = 5{,}500$, with 38.4\% active neurons and 62\% sparsity.
    \item $K_{\mathrm{max}} = 7{,}168$, with 50.0\% active neurons and 50\% sparsity.
    \item $K_{\mathrm{max}} = 9{,}865$, with 68.8\% active neurons and 31\% sparsity.
\end{itemize}

\noindent
\textbf{Methodology.}
We measure sustained autoregressive decoding throughput over 2{,}048 generated
tokens on WikiText-2~\cite{wikitext} test data. We evaluate
model quality with WikiText-2 perplexity and 10-shot normalized HellaSwag
~\cite{zellers2019hellaswagmachinereallyfinish} accuracy.

\begin{figure}[t]
    \centering
    \includegraphics[width=0.95\columnwidth]{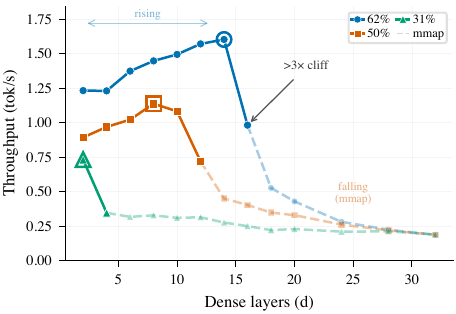}
    \caption{Throughput versus dense prefix length at 11.9\,GiB CUDA-available
    memory. Dashed lines mark configurations whose sparse suffix falls back to
    whole-layer NVMe reads. Circled markers denote the peak $d^*$.}
    \Description{Line chart of decoding throughput against dense prefix length
    for three sparsity levels and an mmap fallback, showing a rising regime, a
    peak, a cliff of more than three times, and a falling regime.}
    \label{fig:ratio_sweep}
\end{figure}

\subsection{Capability in the Model-Exceeds-Memory Regime}
\label{sec:capability}
We evaluate twenty production and research LLM inference systems on a Jetson
AGX Orin under a 14\,GiB memory cap, with 11.9\,GiB available to CUDA, so the
13.5\,GiB Mistral-7B FP16 weights exceed the available resident memory. The
study isolates how runtimes behave when the deployed model exceeds resident
memory, independent of which compression level produced that condition. For
systems that support only the OPT model family, we use OPT-6.7B FP16, whose
12.4\,GiB footprint preserves the same condition.
Table~\ref{tab:capability} summarizes the outcomes. Among the four operational systems, llama.cpp survives through
operating-system demand paging of memory-mapped weights rather than an
explicit sparse execution policy, Ollama inherits this behavior by wrapping
the llama.cpp core, and FlexGen~\cite{flexgen} finds no feasible mixed
placement for its GPU, host, and storage tiers on unified memory, so we report
its all-disk throughput on OPT-6.7B for completeness rather than as a tuned
baseline. At this memory cap, NeuroPrefetcher achieves $9.5{\times}$ the
throughput of llama.cpp, the strongest operational baseline, with the
advantage reaching $12.0{\times}$ across the memory sweep
(Section~\ref{sec:throughput}). The sixteen remaining systems never reach
generation. The six marked \emph{alloc.\ failure} attempt weight transfers
that exceed the unified-memory headroom, and once free memory drops below
roughly 1\,GiB the Jetson kernel cannot satisfy their contiguous allocation
requests. The six marked \emph{unified-memory mismatch} assume separate GPU
and host memory tiers, but on Jetson's single physical pool host offload
creates no new capacity, and the pool is exhausted during initialization.
Candle fails identically despite a compiled systems implementation,
confirming that the issue is architectural rather than framework-specific.
The four marked \emph{no ARM/JetPack release} provide no compatible 64-bit
ARM build for JetPack~6, and version-coupled custom ports are outside the
scope of a deployment-oriented study. The pattern across all classes
indicates a mismatch between traditional offloading designs and
memory-constrained unified-memory edge devices.

\begin{figure}[t]
    \centering
    \includegraphics[width=\columnwidth]{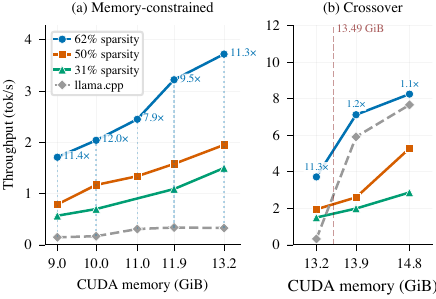}
    \caption{Throughput comparison across memory levels. (a) In the model-exceeds-memory range. (b) Near
    and after the memory crossover, llama.cpp page-fault overhead drops and the
    speedup narrows. The dashed line marks the 13.49\,GiB base model footprint.}
    \label{fig:throughput}
\end{figure}

\subsection{Dense/Sparse Partitioning}
\label{sec:ratio}

Before presenting end-to-end results, we sweep the dense prefix length $d$ to
select the operating point $d^*$, a one-time calibration per hardware and
sparsity setting used in all subsequent experiments. At a fixed 14\,GiB memory
cap, with 11.9\,GiB available to CUDA, we sweep $d$ from 2 to 32 at all three
sparsity levels. The first $d$ MLP layers run dense with resident weights. The
remaining $32{-}d$ MLP layers run sparse using resident active-row buffers and
NVMe delta prefetching. Figure~\ref{fig:ratio_sweep} plots throughput as $d$
increases and exposes the three-regime trade-off that determines $d^*$. The
three regimes appear at all sparsity levels.

\noindent
\textbf{Rising.}
When $d$ is small, most MLP layers run sparse and require NVMe prefetches.
Each additional dense resident layer leaves the sparse prefetch path and no
longer needs storage-to-GPU delta updates, so per-token I/O falls as $d$
increases. The added dense layer also widens the compute window before the
sparse suffix executes, giving the prefetch pipeline more time to fetch
incoming active rows for later layers. At 62\% sparsity, throughput rises from
1.23\,tok/s at $d{=}2$ to 1.60\,tok/s at $d{=}14$.

\noindent
\textbf{Peak.}
The optimum $d^*$ is the largest dense prefix that remains resident while
preserving resident active-row buffers for the sparse suffix. The peak shifts
with sparsity because lower sparsity requires larger active-row buffers. At
62\% sparsity, each sparse layer stores $K_{\mathrm{max}}{=}5{,}500$
active-neuron rows, or 129\,MiB per layer; at 31\% sparsity, each sparse layer
stores $K_{\mathrm{max}}{=}9{,}865$ rows, or 231\,MiB per layer. The larger
buffers leave less room for dense resident MLP weights, so the peak occurs
earlier. At 62\% sparsity, $d^*{=}14$; at 50\% sparsity, $d^*{=}8$; at 31\%
sparsity, $d^*{=}2$.

\noindent
\textbf{Falling.}
\looseness=-1 
Past $d^*$, the dense prefix starves the sparse suffix of resident buffer
space, and layers that lose their active-row buffers fall back to whole-layer
NVMe reads. At $d{=}16$ with 62\% sparsity, the 16 dense MLP layers remain
resident, but two suffix layers lose their buffers and throughput drops from
1.60 to 0.98\,tok/s. Performance continues to fall as resident memory is
increasingly spent on dense placement rather than active-row buffers, reaching
0.19\,tok/s at $d{=}32$, where most MLP layers are served through whole-layer
NVMe reads.

\subsection{End-to-End Throughput}
\label{sec:throughput}

\begin{figure}[t]
    \centering
    \includegraphics[width=\columnwidth]{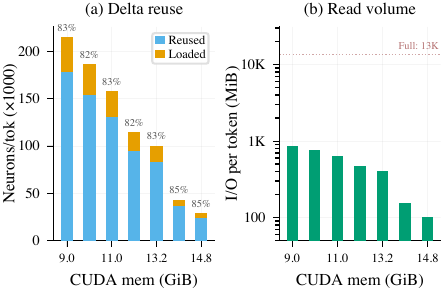}
    \caption{Delta reuse and NVMe read volume at 62\% sparsity.
    (a) Sparse suffix rows reused from the previous token versus incoming rows
    that require I/O. (b) NVMe read volume per token on a log scale. The dashed
    line marks the full 13.49\,GiB model footprint.}
    \label{fig:io_mechanism}
\end{figure}

Using the calibrated dense/sparse split from Section~\ref{sec:ratio}, we compare
NeuroPrefetcher against llama.cpp across seven measured memory levels. The
x-axis reports the available CUDA memory from Table~\ref{tab:memory_levels}. Figure~\ref{fig:throughput} reports throughput at each memory level.

\noindent
\textbf{Model-exceeds-memory.}
From 9.0 to 13.2\,GiB CUDA-available memory, the 13.49\,GiB base Mistral-7B
weight footprint cannot remain fully resident. llama.cpp therefore relies on
memory-mapped demand paging and sustains only 0.2--0.3\,tok/s. NeuroPrefetcher
keeps the calibrated dense prefix resident and executes the suffix through
sparse storage-backed delta reads. At 62\% sparsity, throughput rises from
1.7\,tok/s at 9.0\,GiB to 3.7\,tok/s at 13.2\,GiB. This gives
7.9--12.0$\times$ speedup over llama.cpp across the model-exceeds-memory range.
\looseness=-1
The tightest memory point exposes the buffer-capacity effect from
Section~\ref{sec:ratio}. Lower sparsity uses higher resident active-row
buffers. At 9.0\,GiB, the 50\% and 31\% sparsity settings require larger
buffers than the remaining resident memory can support, even with the smallest
dense prefix, $d{=}2$. Some sparse layers therefore fall back to whole-layer
NVMe reads, pulling their throughput closer to llama.cpp. At 62\% sparsity, the
smaller buffers fit with $d{=}2$, so the sparse suffix remains on
token-to-token delta reads and sustains 1.7\,tok/s.

\noindent
\textbf{Crossover.}
At 13.9 and 14.8\,GiB, llama.cpp throughput rises sharply to 5.9 and
7.7\,tok/s because page-fault-driven weight movement largely disappears.
NeuroPrefetcher also improves with more memory, keeping more dense MLP layers resident and reducing sparse suffix I/O. However, the relative advantage narrows to 1.2$\times$ and 1.1$\times$ because dense
resident execution becomes effective once the model fits. This behavior matches
the design target. NeuroPrefetcher is most useful when the model exceeds
resident memory; after the crossover, sparse storage-backed execution provides
less benefit.

\subsection{I/O Reduction via Delta Reuse}
\label{sec:io}

The throughput gains above come from reducing how much data is read from NVMe
for each generated token. Figure~\ref{fig:io_mechanism} quantifies this effect
at 62\% sparsity. The left panel counts all sparse suffix rows needed per token
and separates rows already resident from incoming active rows that require I/O.
The right panel shows the resulting NVMe read volume.

\noindent
\textbf{Delta reuse.}
Figure~\ref{fig:io_mechanism}(a) shows that 82--85\% of sparse suffix rows are
reused from the previous token across all memory levels. Most sparse suffix demand therefore becomes buffer hits, limiting NVMe traffic to the 15--18\% of newly needed rows. Sparse suffix rows
per token increase as available memory becomes tighter because the calibrated
split keeps fewer MLP layers dense and resident.

\noindent
\textbf{I/O volume.}
Figure~\ref{fig:io_mechanism}(b) shows the system implication of this reuse and
runtime adaptability. The largest and smallest memory settings differ by almost
6\,GiB of available CUDA memory, yet per-token NVMe reads increase only from
103\,MiB at 14.8\,GiB to just under 1\,GiB at 9.0\,GiB. Under tighter memory budgets, NeuroPrefetcher shifts more MLP layers from dense
residency to sparse execution. These additional sparse layers still require
storage traffic, but that traffic is limited to incoming active rows rather than
whole dense MLP layers. This lets NeuroPrefetcher adapt to lower memory budgets
by increasing row-level sparse traffic gradually instead of falling back to much
larger whole-layer NVMe reads.

\subsection{Latency Breakdown}
\label{sec:latency}

\begin{figure}[t]
    \centering
    \includegraphics[width=0.96\columnwidth]{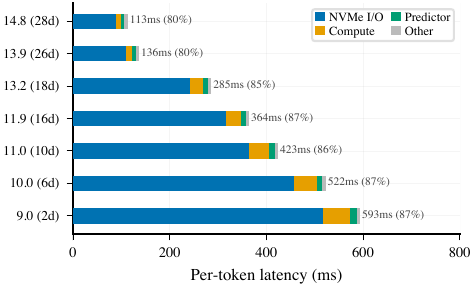}
    \caption{Per-token latency breakdown at 62\% sparsity.}
    \label{fig:latency}
\end{figure}

Figure~\ref{fig:latency} decomposes per-token latency into NVMe I/O, GPU
compute, predictor inference, and remaining overhead. The y-axis reports
available CUDA memory and the calibrated dense prefix length $d$. Across all
memory levels, NVMe I/O accounts for 80--87\% of per-token latency. The result
confirms that the model-exceeds-memory regime is dominated by storage movement,
even after NeuroPrefetcher reduces transfers to incoming active rows.

Compute and prediction remain secondary costs. GPU compute occupies a small
fraction of total latency, so optimizing only the computation would have
limited effect. The predictor is also not a bottleneck. It runs once per token
and remains much smaller than the storage time at every memory level.

Latency falls as available memory increases because the calibrated split keeps
more MLP layers dense and resident. At 9.0\,GiB, only $d{=}2$ MLP layers remain
dense, so most layers are in the sparse suffix and require NVMe row updates.
Total latency is 593\,ms per token. At 14.8\,GiB, $d$ increases to 28, leaving
only four sparse layers, and latency falls to 113\,ms. More resident dense
layers reduce the sparse suffix, while the remaining sparse layers continue to
benefit from token-to-token row reuse.

\subsection{Model Quality}
\label{sec:quality}
\looseness=-1 

Sparse execution is useful only if the selected MLP neurons preserve model
behavior. Quality depends on how closely the selected neurons match the
activation pattern that dense execution would have used. In NeuroPrefetcher,
neuron selection has two stages. Offline clustering converts dense activation
masks into sparse centroid masks, which determines how much activation detail
each centroid preserves. At runtime, the predictor selects one centroid for
each gate and up projection in each sparse layer. Across all sparsity levels
and both models, the predictor selects the correct centroid for roughly 99\%
of layer decisions, so the quality effects measured below stem almost entirely
from the clustering stage rather than from prediction errors. We measure the
combined effect using the same predictor, centroids, and sparse execution path
used by the runtime.

\begin{figure}[t]
    \centering
    \includegraphics[width=0.95\columnwidth]{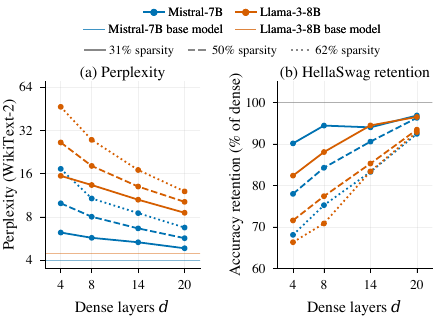}
    \caption{Model quality across dense/sparse configurations for Mistral-7B
and Llama-3-8B. (a) WikiText-2 perplexity, where lower values are
better. (b) HellaSwag accuracy retention relative to each model's dense baseline.}
    \label{fig:quality}
\end{figure}

\noindent
\textbf{Perplexity.}
\looseness=-1 
Figure~\ref{fig:quality}(a) shows WikiText-2 perplexity as the dense prefix
length $d$ increases. For both models and all sparsity levels, perplexity
improves with larger $d$, because each additional dense layer removes one
layer from predictor-guided sparse execution and restores exact dense
computation. The quality cost depends on both sparsity and model. Mistral-7B
is more tolerant of sparse execution across the tested range. At $d{=}20$, its
perplexity remains near the dense baseline for 31\% sparsity and stays below 7
even at 62\% sparsity. Llama-3-8B is more sensitive, especially at high
sparsity and small dense prefixes, with its largest perplexity increase at
$d{=}4$ and 62\% sparsity. Increasing $d$ consistently reduces this gap,
showing that the dense/sparse split provides a direct quality control knob.

\noindent
\textbf{Downstream accuracy.}
Figure~\ref{fig:quality}(b) reports HellaSwag accuracy retention relative to
each model's dense baseline. Retention shows the same trend, improving as $d$
increases and as sparsity decreases. At $d{=}20$, both models recover most of
their dense accuracy across sparsity levels. Mistral-7B retains roughly
95--96\% of dense accuracy, while Llama-3-8B retains roughly 92--93\%
depending on sparsity. These results show a usable operating range rather than
a single fixed configuration. Because the centroids and predictors are
constructed from WikiText-2 activations while HellaSwag differs in both domain
and task format, this retention also indicates that the learned activation
patterns transfer beyond the preparation corpus without per-task retraining.
Increasing $d$ recovers accuracy, while higher sparsity and smaller dense
prefixes support tighter memory budgets. The runtime therefore provides a
controlled tradeoff among memory budget, throughput, and model quality.

\section{Conclusion}
\label{sec:conclusion}
This paper presents NeuroPrefetcher, a storage-backed sparse inference runtime
for the regime where an LLM exceeds resident memory throughout execution.
Instead of treating storage as a reactive overflow tier, NeuroPrefetcher makes
weight movement explicit. Our evaluation shows that this design improves inference on real unified-memory
edge hardware. In the memory-constrained regime, NeuroPrefetcher achieves
7.9--12.0$\times$ speedup over llama.cpp by replacing page-fault-driven
whole-weight movement with application-scheduled sparse-row reads. Delta
reuse keeps 82--85\% of each layer's sparse rows resident across tokens,
so only the changed slice is fetched per token. The capability study shows that this regime is not well served by current
runtimes. NeuroPrefetcher is the fastest distinct operational system in
this setting.
The latency breakdown also shows the remaining bottleneck. Even after delta
prefetching, NVMe I/O dominates per-token latency. Future work should therefore
focus on making the I/O path faster through deeper prefetch pipelines, stronger
read coalescing, quantized sparse-row formats, and better overlap between
storage transfers and GPU execution. The central lesson is that
model-exceeds-memory inference requires runtimes to manage storage traffic at
the granularity of the rows each token actually needs.

\begin{acks}
We thank the anonymous reviewers for their suggestions and feedback. This research was in part supported by US NSF under Grants: SHF-2210744, OAC-2602212, and ERI-2501978.
\end{acks}

\bibliographystyle{ACM-Reference-Format}
\bibliography{sample-base, software}

\end{document}